%% file: main.tex
\documentclass[review]{elsarticle}
\usepackage{mathtools}
\usepackage{amssymb}
\usepackage{subfigure}
\usepackage{caption}
\usepackage{multirow}

\begin{document}

\begin{frontmatter}

\title{A Hierarchical Deep Convolutional Neural Network and Gated Recurrent Unit Framework for Structural Damage Detection}
%HCG: A Hierarchical Deep CNN and GRU Framework for Structural Damage Detection

%\tnotetext[mytitlenote]{Fully documented templates are available in the elsarticle package on \href{http://www.ctan.org/tex-archive/macros/latex/contrib/elsarticle}{CTAN}.}

%% Group authors per affiliation:
\author[address1]{Jianxi Yang}

\author[address1]{Likai Zhang}

\author[address2,corre]{Cen Chen}

\author[address3]{Yangfan Li}

%----------
\author[address1]{Ren Li}
\author[address1]{Guiping Wang}
\author[address1]{Shixin Jiang }
\author[address2,corre]{Zeng Zeng}
%% or include affiliations in footnotes:
%\author[mymainaddress,mysecondaryaddress]{Elsevier Inc}
%\ead[url]{www.elsevier.com}

%\author[mysecondaryaddress]{Global Customer Service\corref{mycorrespondingauthor}}
%\cortext[mycorrespondingauthor]{Corresponding author}
%\ead{support@elsevier.com}

\address[address1]{School of Information Science and Engineering, Chongqing Jiaotong University \\ yjx@cqjtu.edu.cn \\ 622170086019@mails.cqjtu.edu.cn \\ renli@cqjtu.edu.cn \\ wgp@cqjtu.edu.cn \\ shixinjiang@cqjtu.edu.cn}
\address[address2]{Institute for Infocomm Research (I2R), A*STAR \\ chen\_cen, zengz@i2r.a-star.edu.sg}
\address[address3]{College of Computer Scienceand Electronic Engineering, Hunan University \\ yangfanli@hnu.edu.cn}
%\address[address4]{School of Civil Engineering, Chongqing Jiaotong University \\622170086019@mails.cqjtu.edu.cn}
\address[corre]{Corresponding author}

\begin{abstract}
Structural damage detection has become an interdisciplinary area of interest for various engineering fields, while the available damage detection methods are being in the process of adapting machine learning
concepts. Most machine learning based methods heavily depend on extracted ``hand-crafted" features that are manually selected in advance by domain experts and then, fixed. Recently, deep learning has demonstrated
remarkable performance on traditional challenging tasks, such
as image classification, object detection, etc., due to the powerful feature learning
capabilities. 
This breakthrough has inspired researchers to explore deep learning techniques for structural damage detection problems.
However, existing methods have considered either spatial relation (e.g., using convolutional neural network (CNN)) or temporal relation (e.g., using long short term memory network (LSTM)) only. In this work, we propose a novel Hierarchical CNN and Gated recurrent unit (GRU) framework to model both spatial and temporal relations, termed as HCG, for structural damage detection. Specifically, CNN is utilized to model the spatial relations and the short-term temporal dependencies among sensors, while the output features of CNN are fed into the GRU to learn the long-term temporal dependencies jointly.
Extensive experiments on IASC-ASCE structural health monitoring benchmark and 
scale model of three-span continuous rigid frame bridge structure datasets have shown that our proposed HCG outperforms other existing methods for structural damage detection significantly.

\end{abstract}


\end{frontmatter}

%\linenumbers

%\input{abstract}
\input{introduction}
\input{background}
\input{problem_definition.tex}

\input{methodology}

\input{experiment}

\input{conclusions}

\section*{Acknowlegments}
This work is supported by the Science and Technology Planning Project of Yunnan Province (2017IB025), the Science and Technology Research Program of Chongqing Municipal Education Commission of China (KJQN201800705, KJQN201900726), and the Breeding Program of National Natural Science Foundation in Chongqing Jiaotong University (PY201834).

%\section*{References}
%\bibliographystyle{alpha}
%\bibliography{mybibfile}
%\usepackage[numbers, sort]{natbib}
%\bibliographystyle{alpha}
\bibliography{mybibfile.bib}

\end{document}

%% file: introduction.tex
%\linenumbers
\section{Introduction}\label{sec:introduction}
With the rapid development of sensor data acquisition, signal processing, data storage and analysis, Structural Health Monitoring (SHM) system has a wide application prospect for machinery, high-rise buildings, long-span bridges, etc. Many complex structures, such as the Tsing Ma Bridge in Hong Kong, China, the Z24 Bridge in Switzerland, and the Caijia Jialing River Bridge in Chongqing~\cite{zhou2017health}, have deployed SHM systems. A large number of monitoring sensors, such as acceleratormeter, energy, temperature and humidity gauge, and strain gauge, are installed in the key positions of structural buildings, in order to detect abd analyze structural defects. In SHM, structural damage detection based on acceleration vibration signal is one of the most important tasks, which aims to detect damages based on the changes of important structural parameters, such as natural frequency, stiffness, damping ratio and modal vibration mode~\cite{zhouhong2014damage}.

%descirbe the spatio-temporal dependencies

Recent advances in deep learning \cite{chen2017gpu, lecun2015deep} can model the complex nonlinear relationships and
have demonstrated superior performance on a wide range of domains, such as computer vision (CV) , natural language (NLP) processing \cite{dau2020recommendation}, stochastic configuration networks (SCNs)~\cite{wang2017stochasticFA} etc. Stochastic configuration networks~\cite{wang2017stochastic} used widely a lagre-scale data analytical, which can deal with heterogeneous features through a fast decorrelated neuro-ensemble. Convolutional neural networks (CNNs) \cite{chen2018exploiting,han2019convolutional} based models have been successfully utilized to extract spatial features of images which are usually 2D data, and have achieved promising results in CV tasks, such as
image classification \cite{shi20173d}, image segmentation \cite{SeSeNet}, and object detection.
CNNs can extract features because of
two key properties: spatially shared weights and spatial pooling, while recurrent neural networks (RNNs) based methods can generate and
address memories of arbitrary-length sequences of input patterns \cite{Zhang2017A}.
RNN attempts to map from the entire history of previous inputs to target vectors in principle and allows a memory of previous inputs kept in the network’s internal state. RNNs are usually utilized for supervised learning tasks with sequential input data, such as sentiment classification \cite{chen2019gatedsen} and target outputs. GRU is a simple and yet powerful variant of RNNs for sequence modeling tasks due to the gated mechanism \cite{,cho2014properties,chung2014empirical,fanta2020sitgru}. GRU is carefully designed to memorize historical information and fuse current states, new inputs and historical information together in a recurrently gated way.

The successful applications of deep learning have inspired several attempts to address the challenge of structural damage identification.
These work can be mainly classified into three categories, multilayer perceptron (MLP), CNNs, and RNNs based methods, as discussed in the following.
(i) Guo et al., used the sparse coding neural network as the feature extraction model, and the MLP as the classifier for structural damage identification.
(ii) Abdeljaber~\cite{abdeljaber2017real} et al., proposed a structural damage feature extraction and recognition model based on CNNs. Bao~\cite{bao2019computer} et al., proposed a CNN based method which first converted the time series data collected by SHM into an image, and then utilized CNN to learn the features of the converted images.
(iii) Zhao~\cite{zhao2017machine} et al., proposed a RNN based feature extraction method for collected time series data to identify machine conditions. 

These pioneering attempts have presented superior performance for structural damage detection compared with previous
methods based on traditional prediction methods.
However, none of the work considers both spatial relations of different sensors and temporal sequential relations simultaneously.
In addition, RNN-based approaches are slow and difficult to explore the data with very long-term sequential dependencies due to issues
with gradient backpropagation vanishing, while CNN based approaches have high
memory consumption, and do not involve smooth and interpretable
latent representations that can be reused for down-stream tasks.

In this work, we propose a Hierarchical CNN and GRU framework (HCG) that harnesses the capabilities of CNNs and gated recurrent unit (GRU) jointly to capture the complex nonlinear relations
of both space and time for structural damage identification. HCG consists of two levels of models to capture hierarchical levels of features. 
The high-level model uses GRUs to capture the evolving
long-term sequential dependencies of data captured by sensors, while the low-level model is
implemented with CNNs, to capture the short-term sequential dependencies of the data and interactions among sensors.
Experimental results show that HCG has the following
advantages over existing approaches:
\begin{itemize}
  \item {HCG has a significant performance improvement over
existing deep learning models on both IASC-ASCE structural health monitoring benchmark and scale model three-span continuous rigid frame bridge structure datasets;}
  \item {Compared with RNN-based approaches, HCG is 1.5 times
faster in terms of training time;}
  \item {Compared with CNN-based approaches, HCG requires
roughly 10\% data memory usage and allows for easy latent
feature extraction.}
\end{itemize}

The rest of the paper is organized as follows. 
Section~\ref{sec:relatedword} discusses the related work.
Problem definition and analysis are presented in Section~\ref{sec:definition}.
The architecture of HCG is discussed
in Section~\ref{sec:method}.
Performance evaluation of HCG and conclusions are discussed in
Section~\ref{sec:experiment} and~\ref{sec:conclusion}, respectively.

%% file: background.tex
\section{Related Work}\label{sec:relatedword}
In this section, the related developments in structural damage detection domain are mainly illustrated. For many years, research on structural damage detection methods based on high-frequency vibration signals has received extensive attention from academia and industrial communities. These work can be mainly divided into model-driven and data-driven methods~\cite{diez2016clustering}. 

\subsection{Model-driven Method}
Previous model-driven methods utilize mathematical models and physical theorems to discretize the structures. However, these methods have some limitations in the establishment and modification of complex structural models and the simulation of real excitation conditions~\cite{neves2017structural}. Moreover, due to the large-scale structure detection, the natural frequency of different seasons will change greatly. Since the structure is always in the state of unknown excitation, some methods that have achieved good results in the field of mechanical damage detection, such as wavelet transform and HHT, are also subject to certain restrictions. The model-driven methods cannot update the models with
on-line measured data. Therefore, it cannot be applied flexibly. Model modification is mostly a mathematical process and the physical interpretation of the results is not obvious, which demands manual intervention and judgment. Hence, it is difficult to quantitatively identify the states of the structures~\cite{prvsic2017nature}. In recent years, with the rapid development of intelligent algorithms such as statistical machine learning and deep neural networks, data-driven structural damage identification and state analysis methods have become a research hotspot in SHM, which are used to extract damage sensitive indicators and perform pattern recognition directly from structural sensing data.

\subsection{Shallow Models of Data-driven Methods}
In the data-driven domain, shallow learning models, such as support vector machine (SVM), k-nearest neighbor method and random forest have been studied for structural damage detection. For example, Alamdari~\cite{alamdari2019multi} et al., proposed a multi-data fusion method for structural damage detection, which combined the timing response data sets of multiple sensors such as acceleration data, strain gauge data, and environmental data through data tensor and data extraction. SVM was also conducted to perform damage classification.
Chene~\cite{chen2014semi} et al., proposed a classification framework based on semi-supervised learning for structural damage classification.
Carden et al., proposed a statistical classification method based on structural time series response.
Tibaduiza~\cite{tibaduiza2013damage} et al., proposed a damage detection method using principal component analysis (PCA) and self‐organizing maps.
PAC was used to construct the initial baseline model based on the data collected in different test stages.
All these methods are classified as traditional shallow machine learning models. 

\subsection{Deep Models of Data-driven Methods}
Deep learning aims to deeply imitate the data interpretation process of the abstract essential features of human brain, establish a deep network structure similar to the analytical learning mechanism of human brain, and characterize the multi-layer abstract features of the data through a layer-by-layer learning mechanism. The way of feature extraction makes it more suitable for solving practical problems. Therefore, exploring deep learning on structural damage detection has become a hot topic of research~\cite{Pathirage@2018Structural}.
Guo et al., used the sparse coding neural network as the feature extraction model, selected the MLP as the classifier for structural damage identification.
Abdeljaber~\cite{abdeljaber2017real} et al., proposed a structural damage feature extraction and recognition model based on one-dimensional CNN. 
Yu~\cite{yu2019novel} et al., proposed a structural damage identification and localization method based on deep CNN model. 
Abdeljaberet~\cite{abdeljaber20181} et al., proposed an improved CNN classification model and carried out experimental verifications on the IASC-ASCE Benchmark public dataset, which proved that the method requires only a small amount of training data. 
Bao \cite{bao2019computer} et al., proposed an anomaly detection method based on computer vision and deep learning, which converted the time series data collected by SHM into an image, and then used the image as the training set for deep CNNs. 
However, the existing damage detection methods based on deep neural networks have not yet considered the damage feature extraction from the two dimensions of acceleration time series and space correlations among the data. 

%% file: problem_definition.tex
\section{Problem Definition and Analysis}\label{sec:definition}
In this section, we illustrate the problem of structural damage detection and the problem modeling in detail, following the explicit analysis of the problem.

\subsection{Problem Description}
The states of structures can be altered by normal aging due to usage, environment, accidental events, etc. However, the states must remain in the range specified from the design. Structural damage detection aims to give a diagnosis of the ``state" of the constituent materials of a structure and its constituting parts during the life of a structure. The sensors on the structure record the historical states of the structures which can be used for prognosis for the structures, such as evolution of damage, residual life, and so on.

% The state of the structure must remain in the domain specified in the design, although this can be altered by normal aging due to usage, by the action of the environment, and by accidental events. 

%which makes it possible to consider the full history database of the structure, and with the help of %Usage Monitoring, it can also provide a prognosis (evolution of damage, residual life, etc.).

\subsection{Problem Modeling}
\label{sec:modeling}
\begin{figure}[htbp]
    \centering
    \includegraphics[width=12cm]{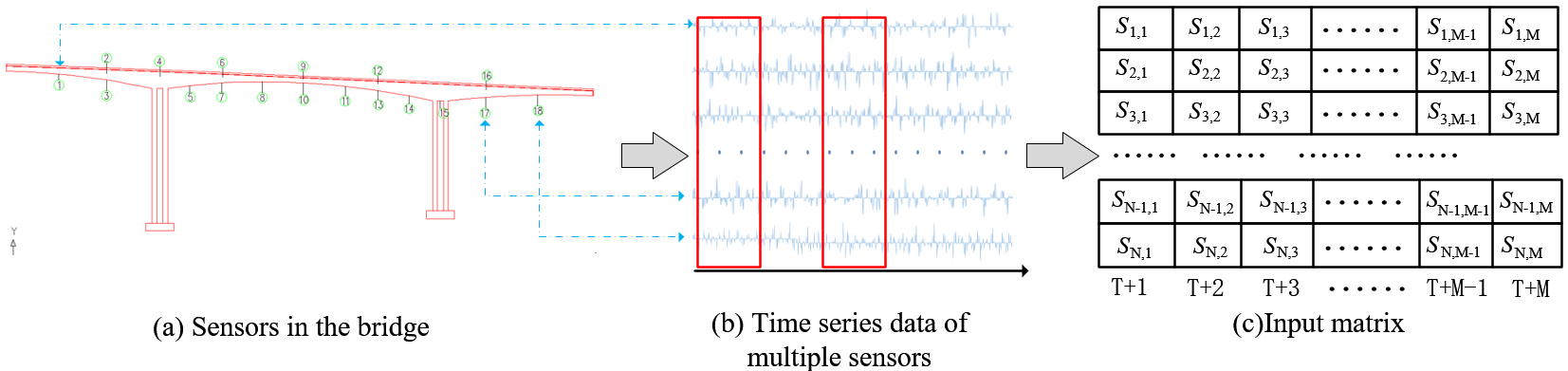}
    \caption{Problem modeling of structural damage detection.}
    \label{problem}
\end{figure}

In this work, we aim to identify the problem of structural damage detection based on historical sensor signals. The modeling of the problem is illustrated in Fig. \ref{problem}.
Sensors are usually deployed on a bridge or other structures to collect data of the states, and each sensor can generate raw time-series data.
The time-series data obtained by all the sensors can be concatenated that form the input matrix of our model, as shown in Fig. \ref{problem}(c). 

Our goal is to estimate the structural damage based on all the sensory data. The structural damage can be divided into several categories, such as Healthy, Damage case 1, Damage case 2, and Damage case 3. Hence, we formulate our problem as a classification problem. 
Formally, given $N$ sensors in total, the $i$-th sensor records a time-series $\boldsymbol{s}^i=\{s^i_1,s^i_2,...,s^i_T\}$, where $s^i_t\in \mathbb{R}$ denotes the sensory value recorded at timestamp $t$, and $T$ denotes the length of the sequence. We suppose that all the sensors are synchronized. Then, we concatenate all the time-series data to form the input matrix $\boldsymbol{X}=\{\boldsymbol{s}^1\circ \boldsymbol{s}^2\circ...\circ \boldsymbol{s}^N\}=\{\boldsymbol{x}_1,\boldsymbol{x}_2,...,\boldsymbol{x}_T\}$, where $\circ$ is the concatenation operation, and $\boldsymbol{x}_t\in \mathbb{R}^N$ represents all the sensor values at timestamp $t$. Then, the output $D$, which represents the estimated damage category, can be formulated as $D=f(\boldsymbol{X})$, where $f$ is the deep neural network we need to design.
% , $N_s$ is the number of categories, $d_i=P(L=i|\boldsymbol{X})$ is the predicted probability for the $i$-th category, $D$ is the predicted label.

\subsection{Problem Analysis}
\label{sec:analysis}
In this section, we first demonstrate that the spatial and temporal dependencies are required to be considered simultaneously for structural damage detection.
Take the SHM for bridges as an example, 
sensors are installed in different positions of a bridge.
Generally, different positions in the structure may have different degrees of forces, and adjacent positions usually withstand similar forces.
Therefore, the data generated by adjacent sensors often have similar patterns and have dependencies with each other.
Meanwhile, the signals collected by the sensor $i$ at the time $t-1$ will affect the sensor's signals at the following time intervals.
We can observe that the data is affected by the spatial and temporal factors simultaneously, which inspires us to design an appropriate model that can learn and extract the spatio-temporal features jointly.
However, by applying either single CNN or GRU model only, the spatio-temporal features cannot be extracted jointly for structural damage detection.

%As discussed above, in order to conduct accurate and robust damage prediction, spatio-temporal %dependencies are required to be considered jointly.
Moreover, due to the notorious gradient
vanishing, GRU and LSTM usually fail to capture very long-term correlation in practice.
In this work, we propose a hierarchical CNN and GRU framework to address these issues, where 
the low-level CNN based model is utilized to extract the spatial and short-term temporal dependencies, while the high-level GRU model is leveraged on to learn the long-term temporal dependencies.

%% file: methodology.tex
\section{Hierarchical CNN and GRU framework (HCG)}\label{sec:method}

\subsection{Architecture of HCG}
\begin{figure}[htbp]
    \centering
    \includegraphics[width=12cm]{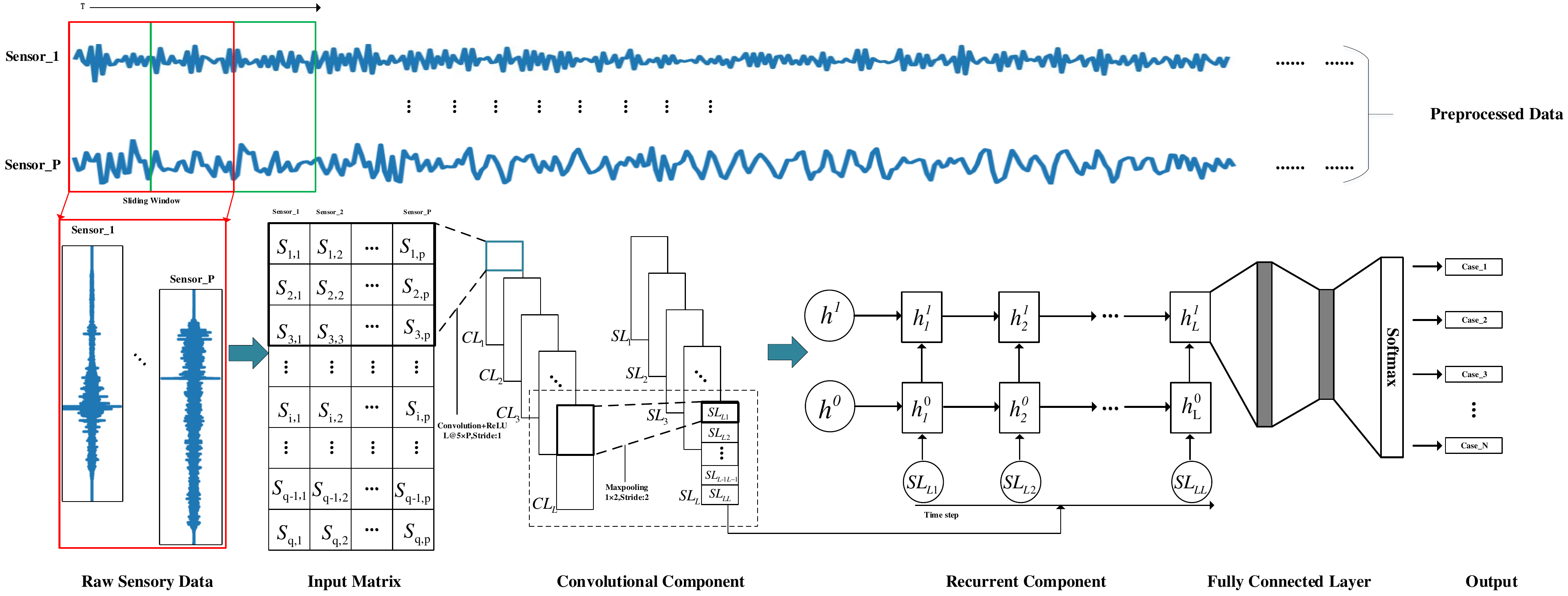}
    \caption{Architecture of our proposed HCG.}
    \label{architecture}
\end{figure}

The architecture of HCG is shown in Fig.~\ref{architecture}.
Our hierarchical model has a
low-level convolutional component that learns from interactions among sensors and the short-term temporal dependencies,
and a high-level recurrent component that handles information across
long-term temporal dependencies. 

The inputs of HCG are the time-series data of multiple sensors, while the outputs are the generated predictions of the structures.
First, the raw sensory time-series data is modeled into the input matrix as discussed in Section \ref{sec:modeling} and then, fed into our proposed convolutional component.
Second, our proposed convolutional component is leveraged on to learn the spatial and short-term temporal features. 
Third, the outputs of the convolutional component are fed into the recurrent component to learn the long-term temporal dependencies.
Finally, a softmax layer is connected with the latent feature vectors generated by the recurrent component to predict the damage state of the structure.

\subsection{Convolutional Component}
As discussed in Section \ref{sec:analysis}, it is crucial to model both the spatial and temporal dependencies in the sensory signals. Convolutional neural network (CNN) is powerful in capturing the spatial correlations and repeating patterns, and has been widely used in image classification, object tracking and video processing \cite{simonyan2014very}, etc. 

\begin{figure}
\centering
\includegraphics[width=10.5cm]{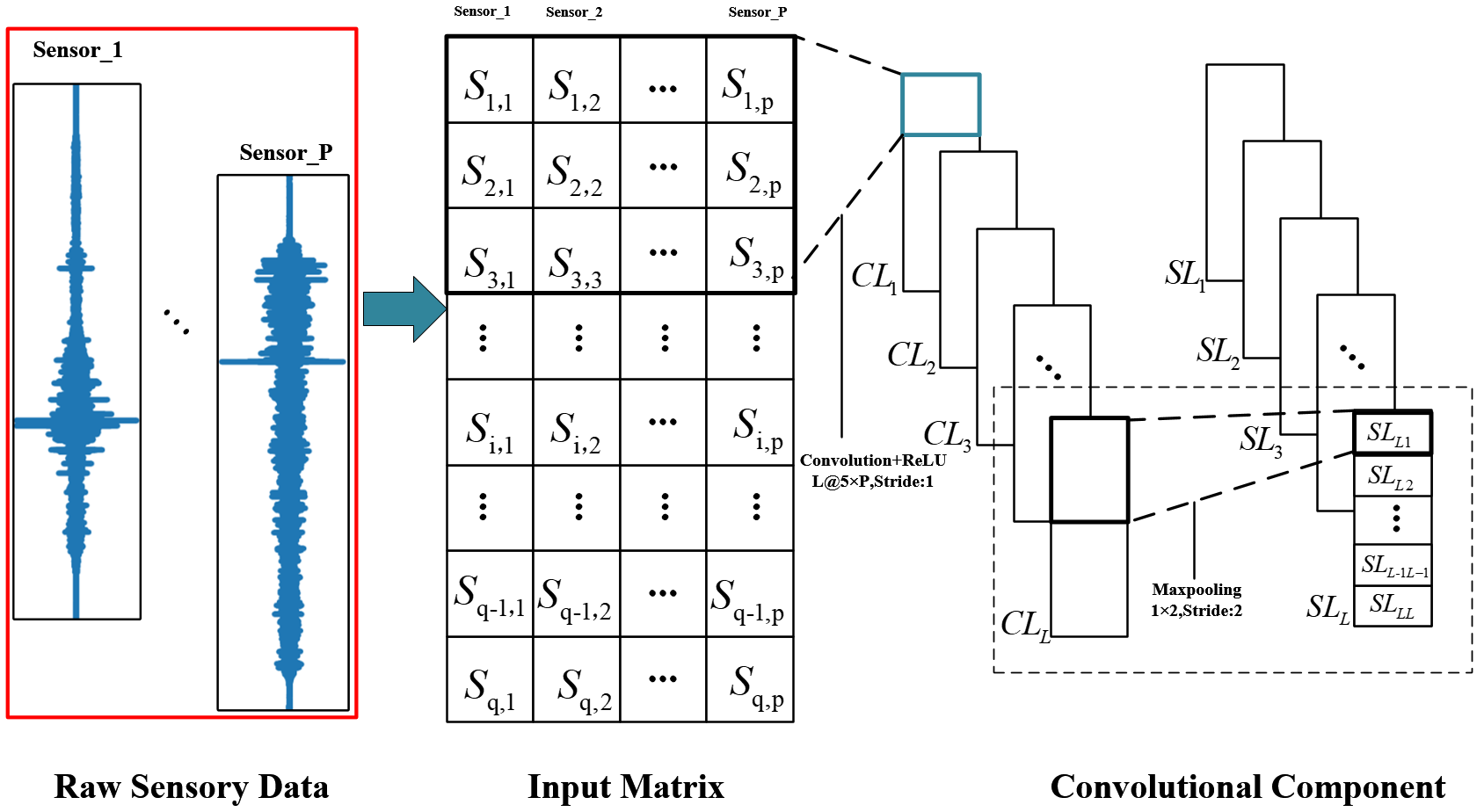}
\caption{Convolutional Component.}
\label{3DCNN}
\end{figure}

A typical convolution layer contains several convolutional filters. Given the input matrix $\boldsymbol{X}$ defined in Section \ref{sec:modeling}, we use these filters to learn the spatial correlations between sensors and the short-term temporal patterns. We let the width of the kernels be the same as the number of sensors so that the kernels are able to capture the spatial correlations among all the sensors. The length of the kernels is relatively short for capturing the short-term temporal patterns. Then, the convolution operation can handle one dimension among the time data as illustrated in Fig. \ref{3DCNN}. The convolution component finally outputs a corresponding sequence where each element is a latent vector and represents the captured patterns at that moment. Zero-padding is used to ensure that the lengths of the input and output are the same. 

Formally, the convolutional layer can be formulated as:
\begin{equation}
F(\boldsymbol{x}_t)=(\boldsymbol{X}*\boldsymbol{f})(t)=\sum_{i=0}^{k-1}\boldsymbol{f}_j\cdot\boldsymbol{x}_{t-j}  ,\boldsymbol{x}_{\leq 0} \coloneqq 0,
\end{equation}
\begin{equation}
\boldsymbol{O}=\{F(\boldsymbol{x}_1),F(\boldsymbol{x}_{2}),...,F(\boldsymbol{x}_{T})\},
\end{equation}
\begin{equation}
{\boldsymbol{Y}}=RELU(\boldsymbol{O}),
\end{equation}
where $\boldsymbol{x}_t \in\mathbb{R}^{N}$ denotes the values of all the sensors at times $t$, $\boldsymbol{f}\in \mathbb{R}^{N\times k}$ denotes a convolution kernel with size $k$, the $RELU$ function $RELU(x)=max(0,x)$ is the activation function, and ${\boldsymbol{Y}}=\{\boldsymbol{y}_1,\boldsymbol{y}_2,...,\boldsymbol{y}_T\}$ denotes the output sequence of our convolutional component. Each element $\boldsymbol{y}_t\in \mathbb{R}^d$, where $d$ is the number of the kernels, denotes the latent representation of the spatial and short-term features at time $t$. 

\subsection{Recurrent Component}
Recurrent neural networks (RNNs) have recently demonstrated the promising results in many machine learning tasks, especially when input and/or output are a sequence of variables.
GRU is a simple and yet powerful variant of RNNs for time series prediction due to the gated mechanism \cite{chen2019gated}. GRUs are carefully designed to memorize historical information and fuse current states, new inputs and historical information together in a recurrently gated way.

\begin{figure}
\centering
\includegraphics[width=8.0cm]{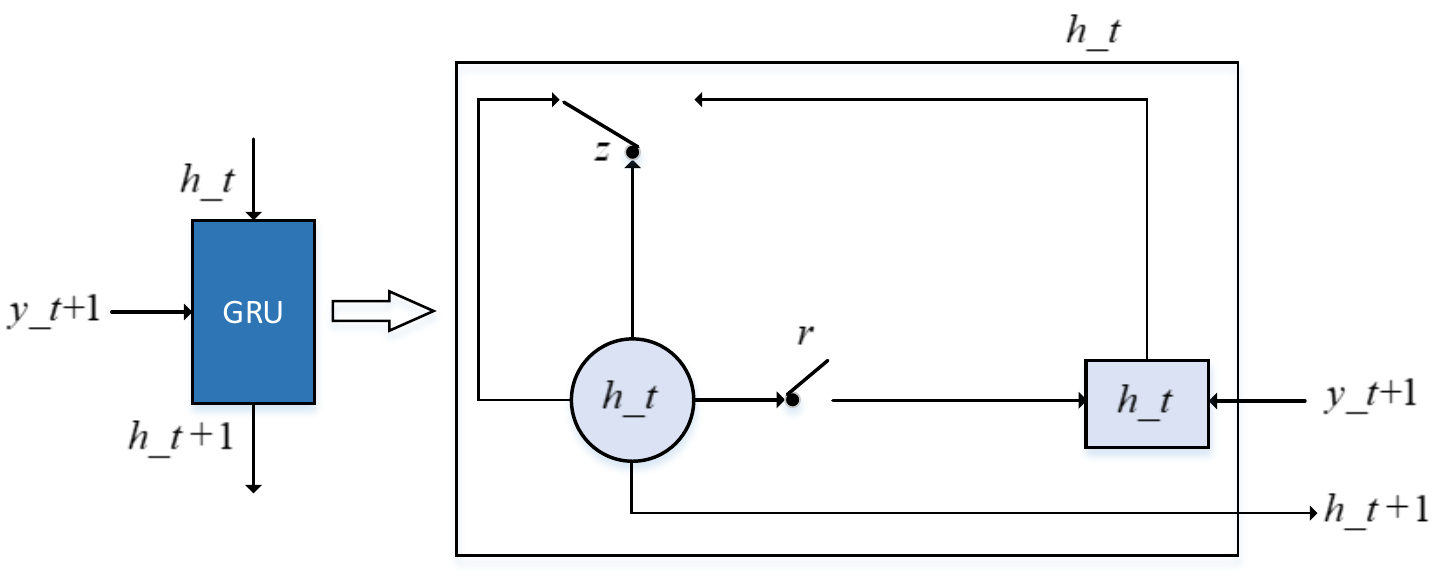}  %{architecture.eps}
\caption{Illustration of Gated Recurrent Unit (GRU) at the iteration step $t$. 
}
\label{gruori}
\end{figure}

The outputs generated by the convolutional components are fed into the GRU to extract the long-term temporal dependencies.
The computing process of the GRU unit at time $t+1$ can be shown in Fig. \ref{gruori} and formulated in Equ. \ref{equ:gnn}:
\begin{equation}
\label{equ:gnn}
\begin{aligned}
r_{t+1} = \sigma(\Theta_r [y_{t+1}, h_{t}] + b_r), \\
u_{t+1} = \sigma(\Theta_u [y_{t+1}, h_{t}] + b_u), \\
c_{t+1} = tanh(\Theta_c [y_{t+1}, (r_{t+1} \odot h_{t})] + b_c), \\
h_{t+1} = u_{t+1} \odot h_{t} + (1-u_{t+1}) \odot c_{t+1}, \\
\end{aligned}
\end{equation}
where $h_{t}$ is the hidden state of a GRU generated at the iteration step $t$ and is the original hidden states for the iteration step $t+1$, $y_{t+1}$ is the hidden features generated by the convolutional component, $h_{t+1}$ is the generated hidden state of a GRU,  $r_{t+1}$ and $u_{t+1}$ are the reset gate and update gate at the time $t+1$, respectively, $\Theta_r, \Theta_u$ and $\Theta_c$ are the learned parameters of filters, and $\odot$ is the element-wise multiplication of tensors.

By recurrently connecting the GRU cells, our recurrent component can process complex sequence data. Then, we use the hidden state at the last timestamp of the top-layer GRU to predict the damage category. We use a fully connected network and a softmax layer to generate the final output of HCG. Formally, the predicted category $D$ is formulated as:
% and the full connection layer of the last sequential state is selected for classification through Softmax. The results obtained by Softmax represent the probability that the input data is divided into each category. In the case of a K classifier, the output is a K-dimensional vector (the sum of the elements in the vector is 1). The Softmax layer is formulated as:
%
\begin{equation}
\label{equ:fc}
\boldsymbol{y}^f=(\boldsymbol{h}_T^{N_l})^{\mathsf{T}}\boldsymbol{W}+\boldsymbol{b},
\end{equation}
\begin{equation}
\begin{aligned}
\label{equ:softmax}
d_i=Softmax(\boldsymbol{y}^f)_i= \frac{{{e^{{y^f_i}}}}}{{\sum\nolimits_{j = 1}^{N_c} {{e^{{y^f_j}}}} }},
\end{aligned}
\end{equation}
% Where ${d_i}$ is the estimated probability of the damage category $i$, $N_c$ denotes the number of categories. Then the final output is
\begin{equation}
D=\arg\max _{i\in 1,...,N_c}d_i,
\end{equation}
% which represents the estimated probabilities for all the categories.
where $\boldsymbol{y}^f\in \mathbb{R}^{N_c}$ is the output of the fully connected layer, $N_l$ is the number of GRU layers, $N_c$ denotes the number of categories, $\boldsymbol{W}$ and $\boldsymbol{b}$ are trainable parameters, $\boldsymbol{h}^i_t$ denotes the hidden state of the $i$-th GRU layer at timestamp $t$, and $d_i=P(D=i|\boldsymbol{X})$ is the predicted probability for the $i$-th category. 

% represents the input, ${l_j}$ is the activation function, and ${W_{ji}}$ is the weight matrix, so the predicted label 11 formula (5) is shown.
% \begin{equation}
% \begin{aligned}
% \label{equ:ymax}
% \hat y = \arg \mathop {\max }\limits_i {p_i}
% \end{aligned}
% \end{equation}

\subsection{Loss Function}
In the training process, we adopt mean squared error (L2-loss) as the objective function, the corresponding optimization objective is formulated as:
\begin{equation}
\min  _{\boldsymbol{\Theta}} \sum _{j=0}^M \sum _{i=0} ^{G}(\boldsymbol{d} _{t,i}^j-\hat{\boldsymbol{d}} _{t,i}^j)^{2},
\label{equ:weightedfusion}
\end{equation}
where $\boldsymbol{\Theta}$ denotes the parameter set of our model, $G$ is the number of training samples, $\boldsymbol{d}$ is the ground truth of the damage state, and $\hat{\boldsymbol{d}}$ is the predicted class of our model.

%% file: experiment.tex
\section{Performance Evaluation and Discussions}\label{sec:experiment}
In this section, we evaluate our proposed HCG on two datasets including a Three-span Continuous Rigid Frame Bridge dataset (TCRF Bridge Dataset) and the phase I IASC-ASCE Structural Health Monitoring Benchmark dataset (IASC-ASCE Benchmark Dataset). First, the details of the datasets will be presented. Then, the experimental settings and implementation details will be illustrated. After that, the experimental results compared with other baselines will be presented and discussed.

\subsection{Datasets}
In this section, two popular datasets in the domain shall be described in detail as follows.

\subsubsection{TCRF Bridge Dataset}
\begin{figure}[htbp]
  \centering
  \subfigure{
    \includegraphics[width=8.0cm]{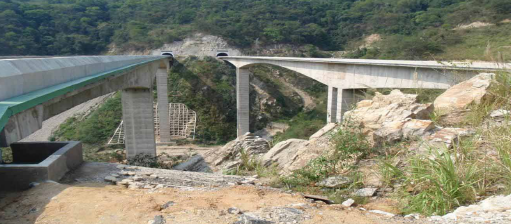}}
  \caption{Three-span Continuous Rigid Frame bridge (TCRF Bridge).}
  \label{heiChongGOU}
\end{figure}

The real Three-span Continuous Rigid Frame Bridge (TCRF Bridge) structure is shown in Fig. \ref{heiChongGOU}, in which the main span is 98m + 180m + 98m and the total length is 377.30m. This bridge adopts the single box and single room structure. The roof width of the box girder is 12.5m. The width of the bottom plate of the box is 6.5m.

\begin{figure}[h]
  \centering
  \subfigure{
    \includegraphics[width=5.8cm]{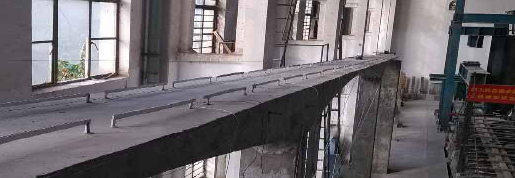}}
  \hspace{0.02cm} 
  \subfigure{
    \includegraphics[width=5.8cm]{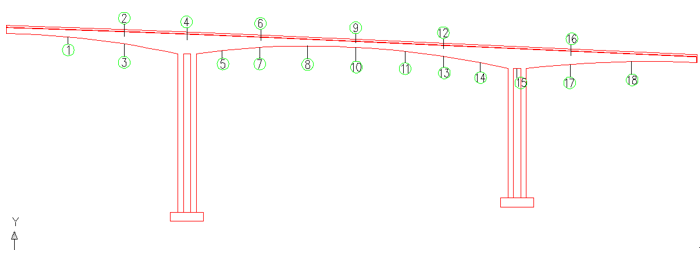}}
  \caption{Sensor placement on the scale model of TCRF Bridge.}
  \label{suoChi}
\end{figure}

In our experiments, we use a scale model of the bridge where the main bridge, bridge pier and bridge abutment are constructed following the same scaling ratio of $20:1$. Then, the stiffness degradation of the bridge structure is simulated by applying the concentrated force in the span of the continuous rigid frame bridge to make the floor crack. In order to monitor the changes of the structural state degradation, $18$ acceleration sensors have been installed on the scale model, including $12$ vertical measuring points at the bottom of the beam and $6$ on the web horizontal measuring points. We tow the trolley on the bridge deck to simulating the dynamic load process, then the acceleration can be monitored by the sensors, as shown in Fig. \ref{suoChi}.

We apply concentrated force on the main span of the scale model that leads to cracks in middle floor of main span. We use these cracks to represent the structural damages. The degree of damage depends on the strength of the concentrated force. In our experiment settings, we have $4$ kinds of structural damage states, which are shown in Table \ref{table:suochi}. 

\begin{table}[ht]
    \centering
    \caption{Structural damage states.}
    \begin{tabular}{c|p{10cm}}
    \hline  
States & Descriptions  \\  
    \hline 
DC0 & No damage in the bridge structure.  \\ 
DC1 & One crack in the bridge.  \\ 
DC2 & Two cracks in the bridge.  \\ 
DC3 & Two larger cracks. \\ 
    \hline  
    \end{tabular}
    \label{table:suochi}
\end{table}  
 
\begin{figure}[htbp]
  \centering
  \subfigure{
    \includegraphics[width=5.8cm]{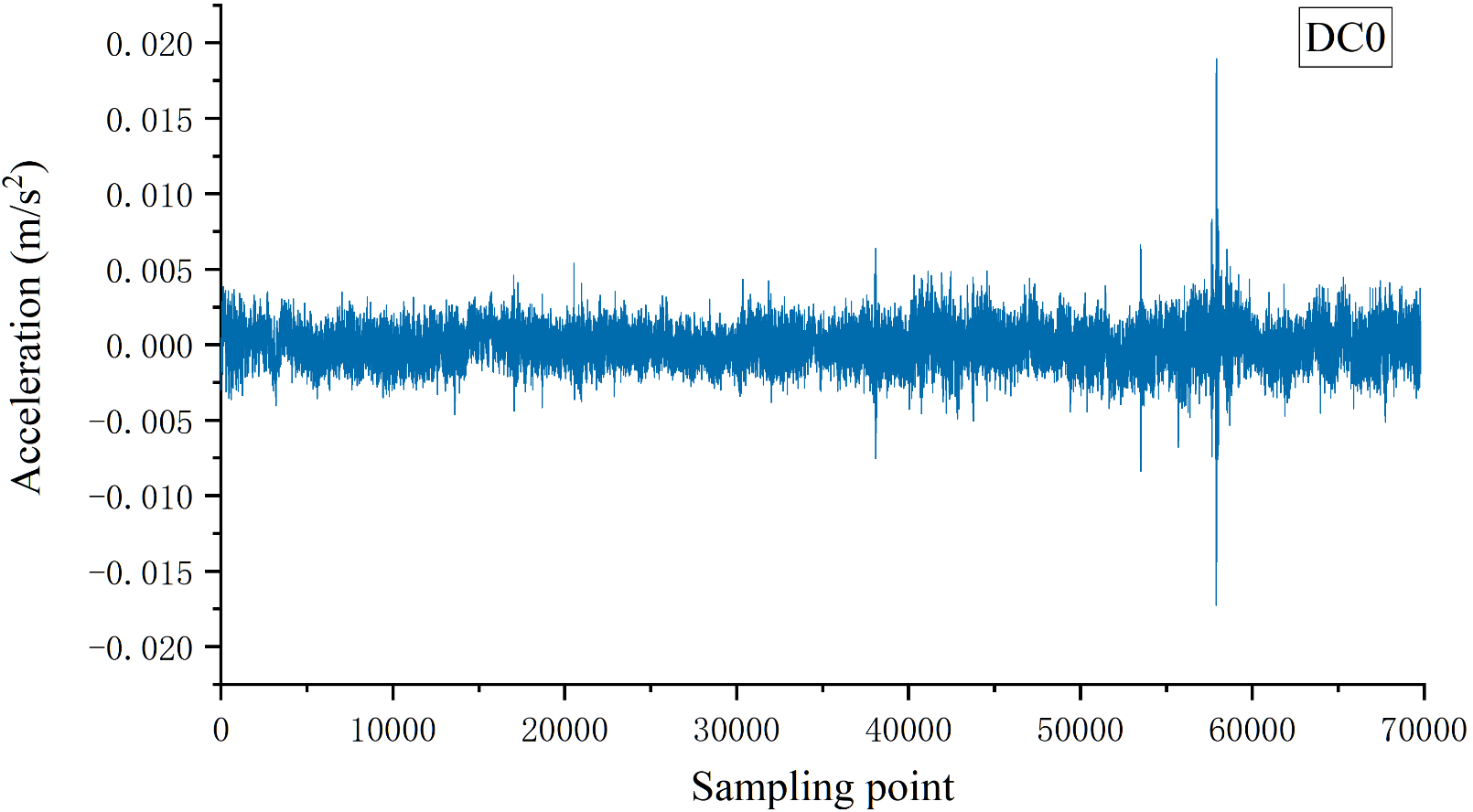}}
  \subfigure{
    \includegraphics[width=5.8cm]{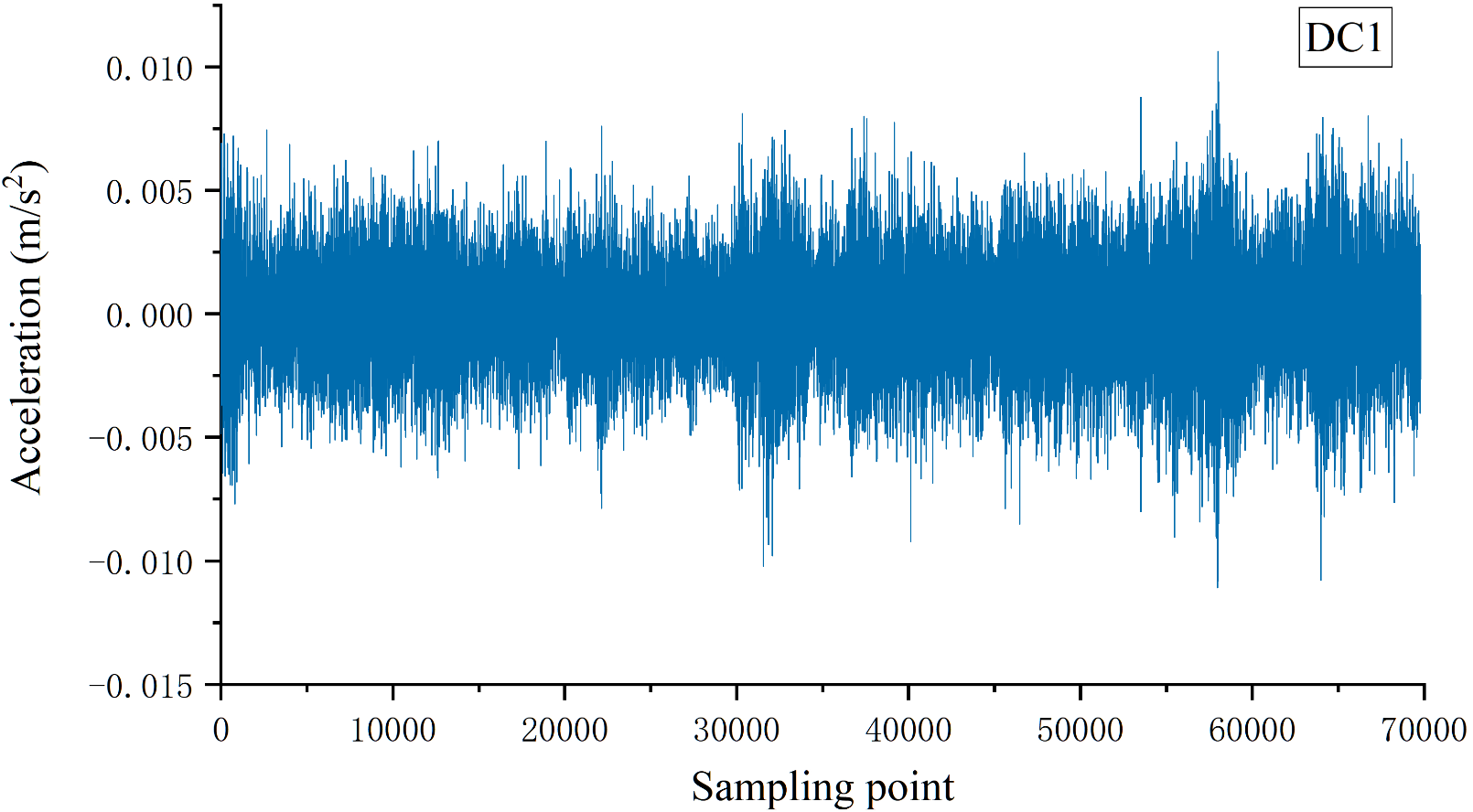}}
  \subfigure{
    \includegraphics[width=5.8cm]{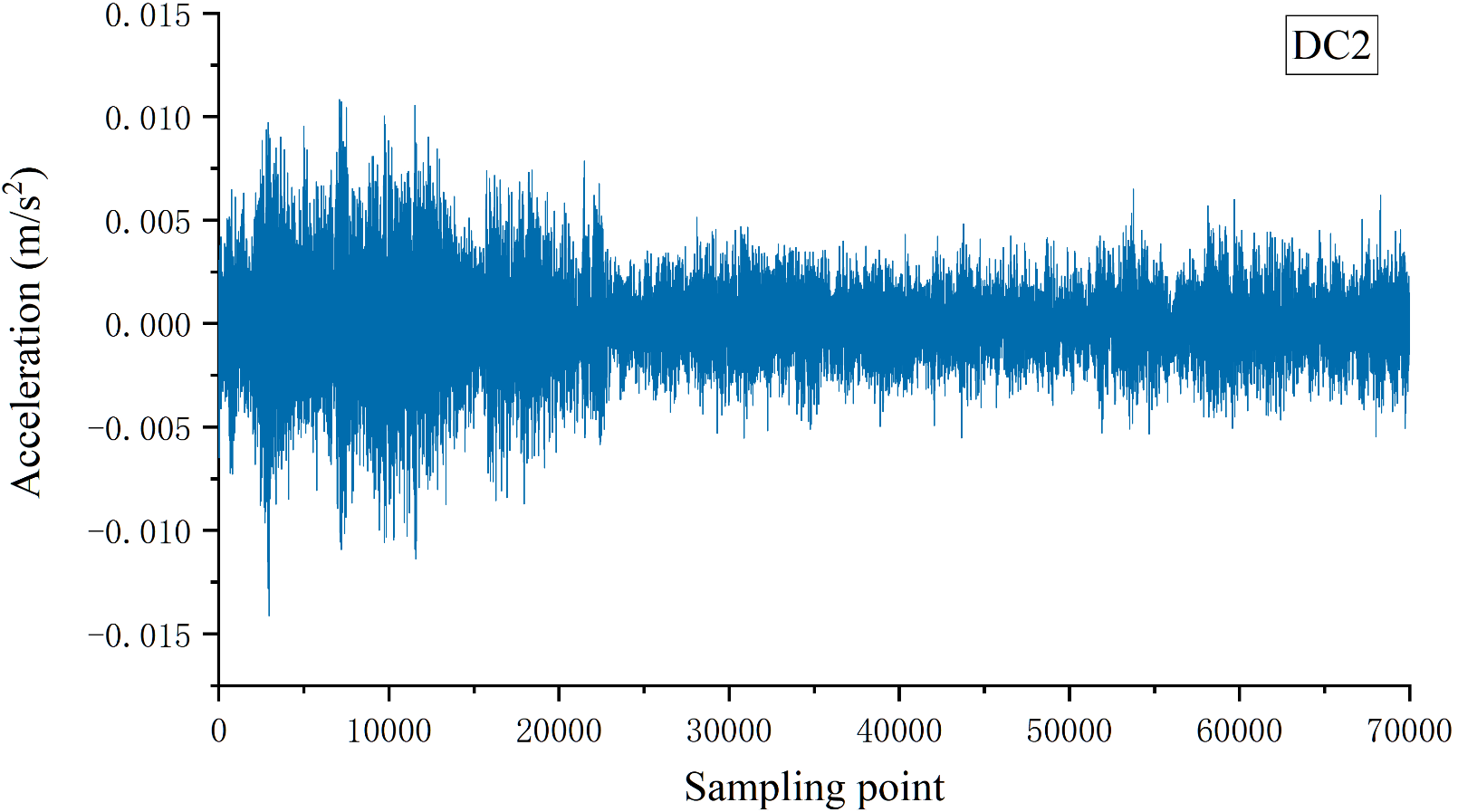}}
  \subfigure{
    \includegraphics[width=5.8cm]{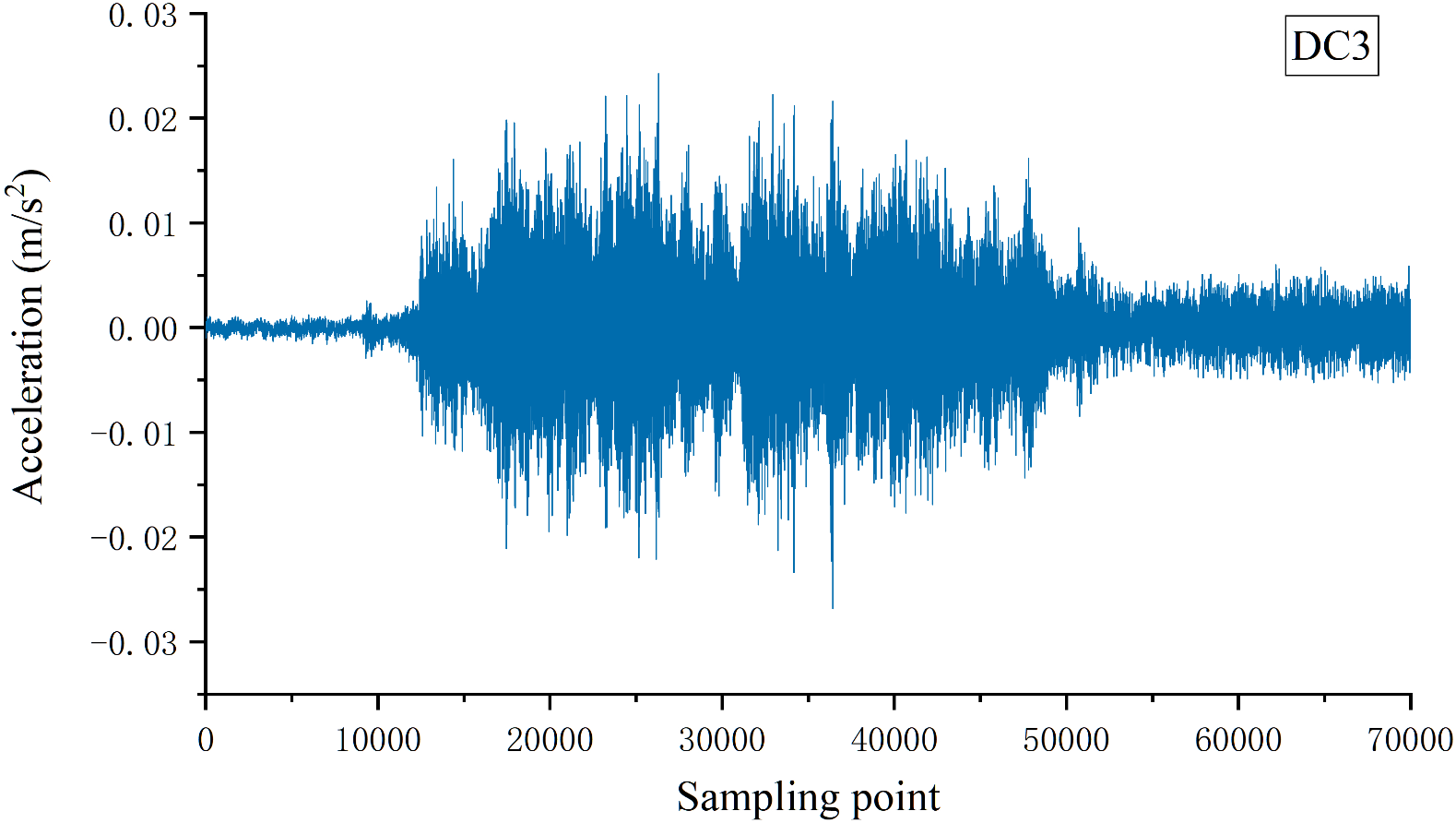}}
  \caption{Different acceleration curves of a sensor for different damage states when a car is passing.}
  \label{Monitoring}
\end{figure}

When a car passes through the bridge deck, the acceleration signals of each sensor are collected with the sampling frequency of $8$Khz. The signals are quite different for different damage states. For example, in the case of different structural damage, the curve of acceleration at the second measuring point is shown in Fig. \ref{Monitoring}. The figure shows that the sensory data keep floating around zero, which accords with the stable time series. It can be observed that the response data are obviously different among the different damage states. The HCG model is used to learn the spatial and temporal features of these sensory data.

% The phase I IASC-ASCE Structural Health Monitoring Benchmark dataset
% (IASC-ASCE Benchmark Dataset) is proposed in \cite{johnson2004phase}. It is a simulated structure and widely used for structural damage classification. 
\subsubsection{IASC-ASCE Benchmark Dataset}
\begin{figure}[htbp]
  \centering
  \subfigure{
    \includegraphics[scale=0.3]{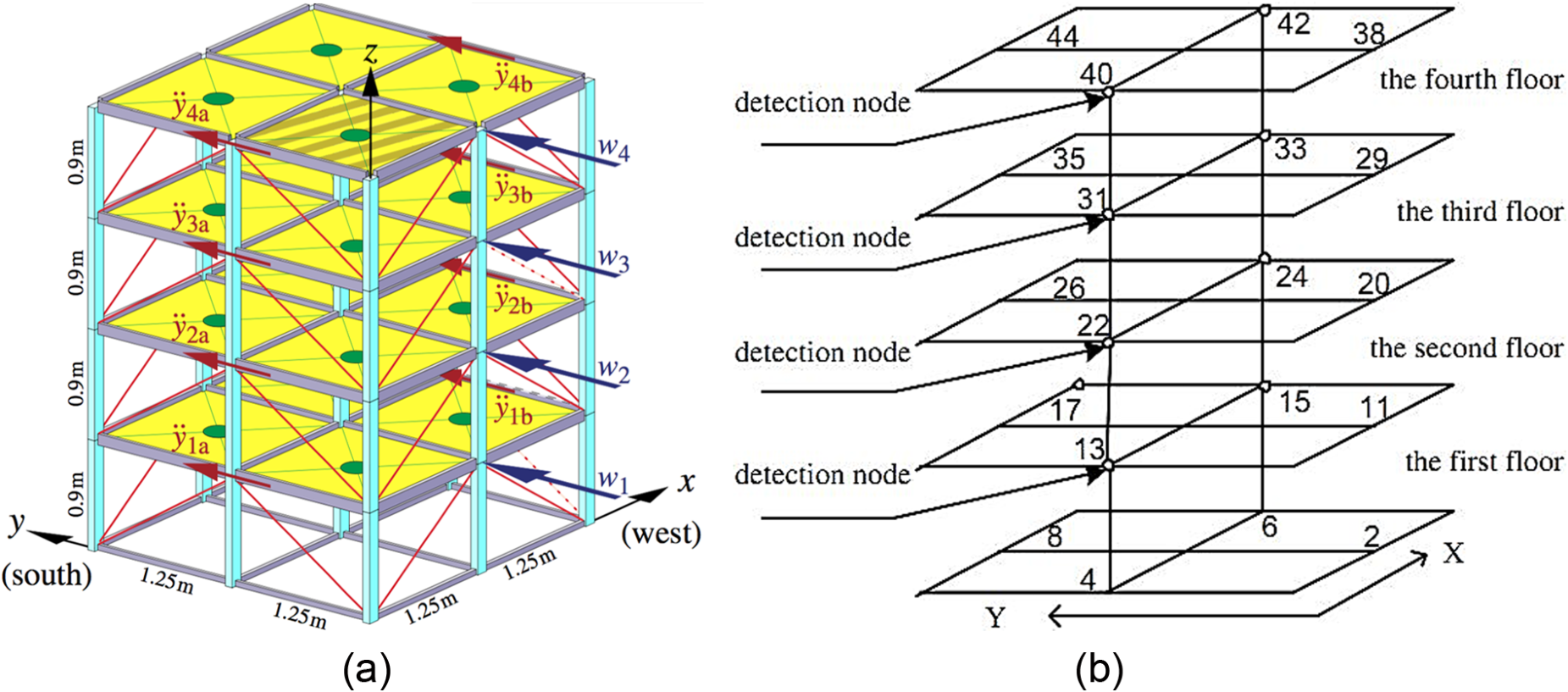}}
\caption{IASC-ASCE Benchmark Dataset: (a) stucture diagram. (b) distribution of the sensor nodes.}
\label{fig:benchmark}
\end{figure}

The phase I IASC-ASCE Structural Health Monitoring Benchmark dataset is a simulated structure and widely used for structural damage classification. 
The primary purpose of the IASC-ASCE benchmarks is to offer a common platform for numerous researchers to apply various SHM methods to an identical structure and compare their performance. The benchmarks comprise two phases, e.g., Phase I and Phase II, each with simulated and experimental benchmarks. In this study, the phase I IASC-ASCE Structural Health Monitoring Benchmark dataset (IASC-ASCE Benchmark Dataset) is proposed in. The simulated structure is shown in Fig.~\ref{fig:benchmark}.

For the IASC-ASCE Benchmark structure, the degree of freedom (DOF) is 120, and the mass distribution is symmetric.
The damage states are shown in Table \ref{tab:benchmark}. Sensors are installed on each floor of the middle column along both sides. There are $16$ sensors in total. We use shaker on roof as excitation signals. The response acceleration signals are gathered from sensors on column $4$ of each floor. The sampling frequency is $250$Hz and the length of the data is $20,000$. Then, the collected sensory data is fed to HCG as the inputs.

\begin{table}[ht]
    \centering
    \caption{IASC-ASCE Benchmark Dataset damage states.}
    \label{tab:benchmark}
    \begin{tabular}{c|p{10cm}}
    \hline  
States & Descriptions  \\  
    \hline  
DC0 & Without damage.  \\ 
DC1 & Remove all inclined supports from the first floor.\\ 
DC2 & All braces in $1$-st and $3$-rd stories remove.\\ 
DC3 & Remove an oblique support from the first floor.\\ 
DC4 & Remove one oblique support from the first floor and one from the third floor.\\ 
DC5 & Damage $4$ + relaxation $18$ element (first floor beam element) to the left.\\ 
DC6 & The area of a certain inclined support on the first floor is reduced by $1/3$. \\ 
    \hline  
    \end{tabular}
\label{table:Benchmark}
\end{table}  

\subsection{Experimental Setup}
\subsubsection{Experiment Settings}
The two datasets are divided into training set ($60\%$), validation set ($20\%$) and test set ($20\%$). 
Keras is used to build our models. We train and evaluate all the methods on a server with $4$ Tesla P100 GPUs and $8$ E5-2620V4 CPUs. 

\subsubsection{Evaluation metrics}
We utilize $4$ general evaluation metrics: Accuracy, Precision, Recall and F1 value to evaluate HCG and other compared baselines.

\begin{equation}
\label{equ:metric}
% Accuracy = {{TP + TN} \mathord{\left/
%  {\vphantom {{TP + TN} {All{\rm{ }}samples}}} \right.
%  \kern-\nulldelimiterspace} {All{\rm{ }}samples}}
Accuracy=\frac{TP+TN}{TP+FP+TN+FN},
\end{equation}
\begin{equation}
    % MicroR = \frac{{\overline {TP} }}{{\overline {TP}  \times \overline {FP} }}\\
Precision=\frac{TP}{TP+FP},
\end{equation}

\begin{equation}
    % MicroR = \frac{{\overline {TP} }}{{\overline {TP}  \times \overline {FP} }}\\
Recall=\frac{TP}{TP+FN},
\end{equation}

\begin{equation}
    % MicroF1 = \frac{{2 \times MicroP \times MicroR}}{{MicroP \times MicroR}}\\
F_1=2\times\frac{Precision\times Recall}{Precision+Recall},
\end{equation}
where $TP$, $FP$, $TN$ and $FN$ denotes true positives, false negatives, false positives and true negatives, respectively.
% where $\overline {TP}$ each class is predicted to be a positive sample with positive prediction, and its average is evaluated. where $\overline {FP}$ each class is predicted to be a positive sample with positive prediction, and its average is evaluated

% Each class is predicted to be a positive positive sample, and its average is evaluated

\subsubsection{Compared Baselines}
We compare HCG with a variety of baseline methods, which are summarized as follows. To ensure fair comparison, for all deep learning based model, we adjust the layer number and hidden units number such that all the models have very similar number of trainable parameters. All the deep learning based models are trained with Adam optimizer, with learning rate $0.001$ and batch size $64$. 

\begin{itemize}
\item DNN: 
We use a $3$-layer fully connected neural network with softmax activation. Each layer has $512$, $256$ and $128$ neurons, respectively.

\item CNN \cite{abdeljaber2017real}: We adopt a $3$-layer CNN where the number of kernels is $32$, $64$, and $32$, respectively. We use the $5\times5$ convolutional kernels for all the layers. 
%CNN  32个过滤层、每层的卷积核的大小为（5,5,）在通过一个maxpool（2,2,）

\item GRU: We adopt a GRU model by stacking $3$ layers of GRU cells described in Equ. \ref{equ:gnn}, each with $64$ dimensional hidden state. 

\item LSTM: We improve the method in \cite{zhao2017machine} by adopting a LSTM model by stacking $3$ layers of LSTM cells, each with $64$ dimensional hidden state.

\item HCG: For the convolutional component, the size of the convolution kernel is ($5$, N) (N denotes the number of sensors) and the number of filters is $64$, and $2$ layers are stacked. For the recurrent component, a $2$-layer GRU is connected where each layer with $128$ dimensional hidden state. Then, two full connections having $256$ and $128$ neurons with softmax activation generate the final prediction.
\end{itemize}

\subsection{Results of the TCRF Bridge Dataset}	
Table \ref{tab:hei-acc_loss} summarizes the experimental results of our proposed HCG and other baselines on the TCRF Bridge Dataset. The Dataset adopts the standard deviation. Then we run all the baselines $10$ times and present the average results.
HCG obtains the better results compared with single CNN and GRU model.
It shows the effectiveness of considering both the spatial and temporal dependencies together.
HCG also outperforms DNN and LSTM based models.
Overall, our proposed HCG outperforms other baselines including DNN, CNN, LSTM and GRU on all the evaluation metrics.

\begin{figure}[htbp]
  \centering
  \subfigure{
    \includegraphics[width=5.8cm]{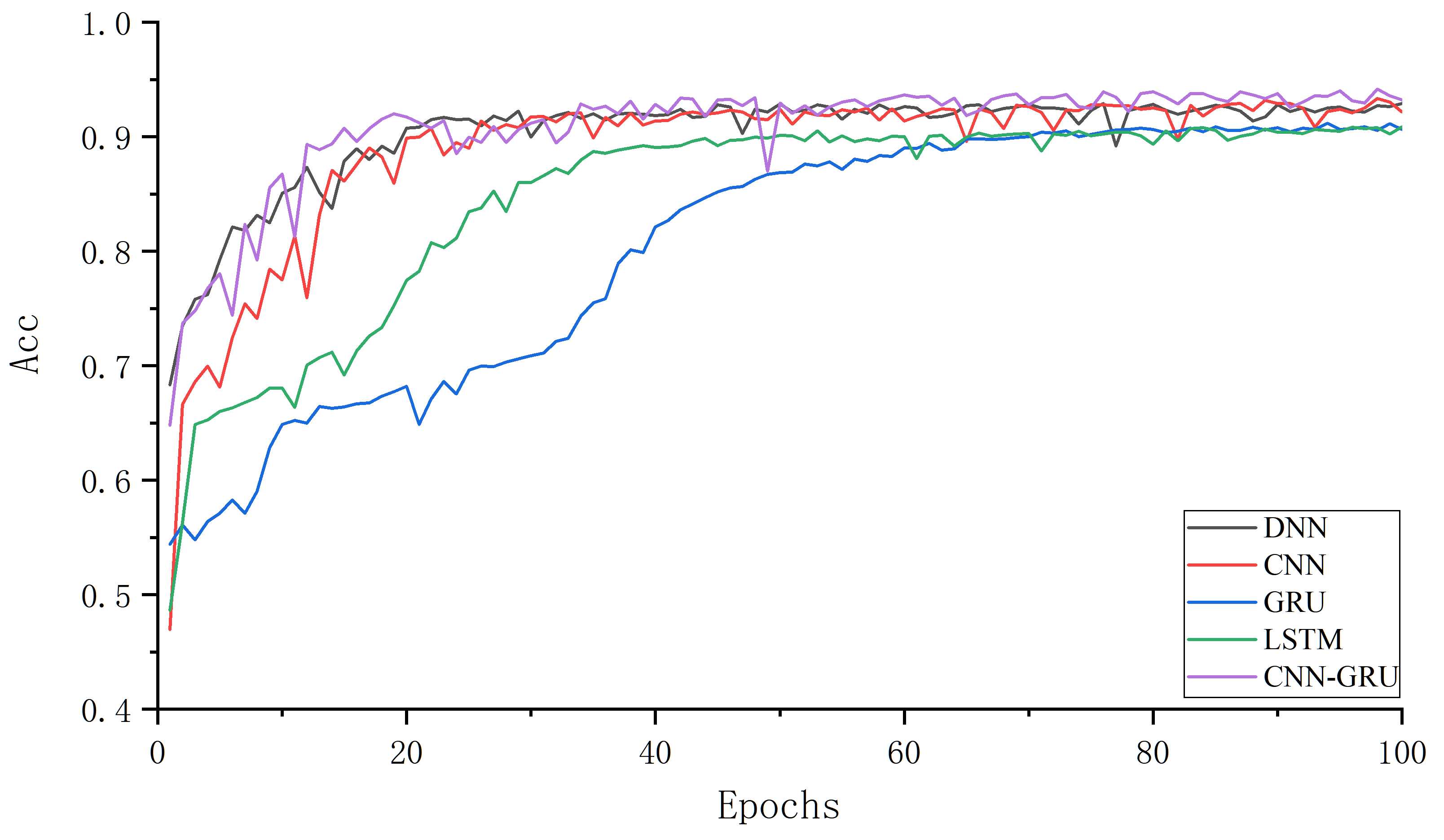}}
  \subfigure{
    \includegraphics[width=5.8cm]{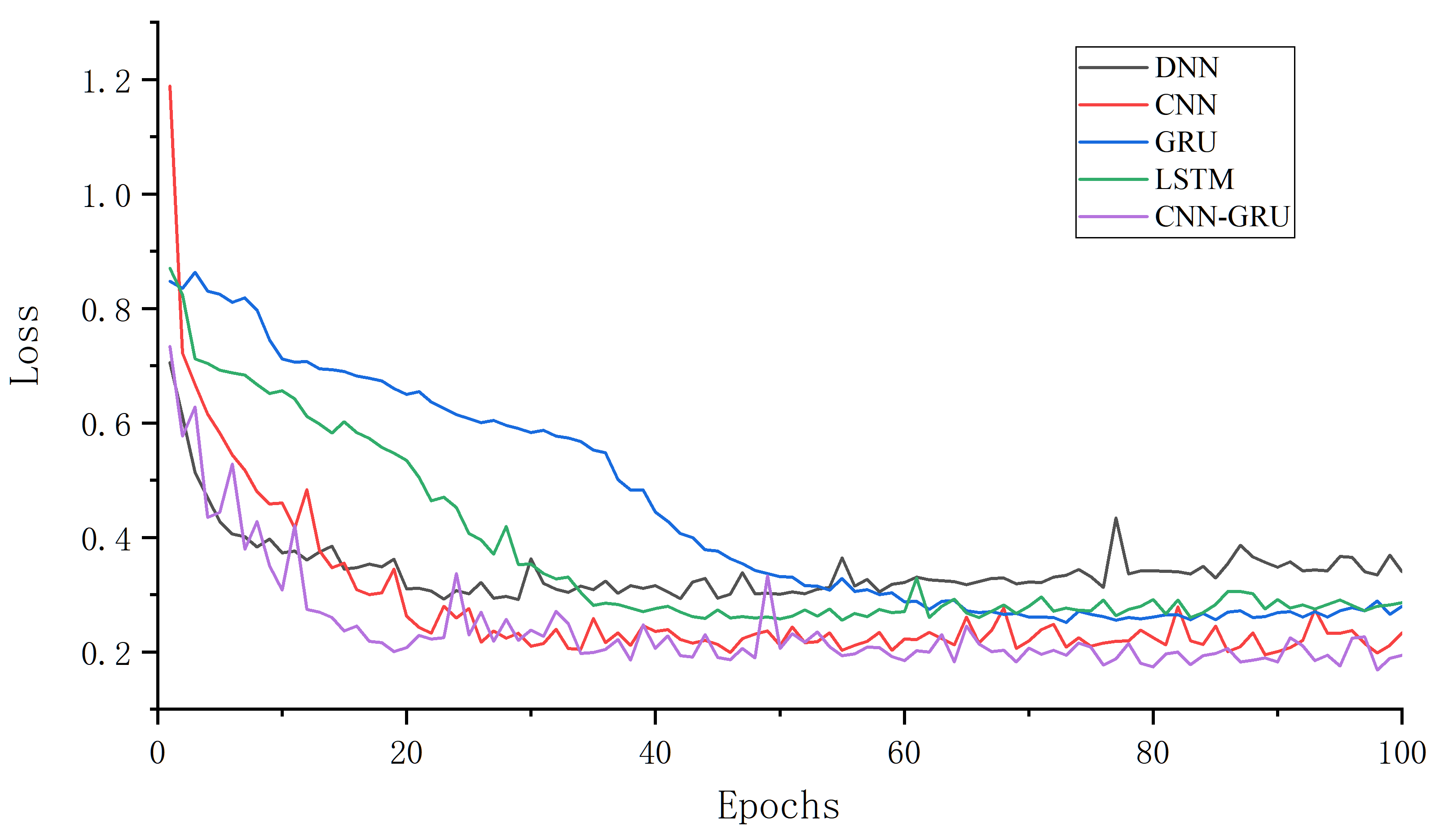}}
\caption{Loss curve and accuracy curve in the training process for the TCRF Bridge Dataset. (a) The accuracy curve. (b) The loss curve.}
\label{fig:hei-result}
\end{figure}

%In \ref{tab:hei-acc_loss}, HCG outperforms other competitors and in Fig.\ref{fig:hei-result}, 
Fig. \ref{fig:hei-result} illustrates the loss curve and accuracy curve in the training process for the deep-learning based baselines and our HCG.
We can clearly observe that our proposed HCG converges more quickly than other competitors which demonstrate HCG's priority as HCG can capture both the spatial and temporal dependencies in the sensory data.

\begin{table}[h]
    \centering
    \caption{Results of the TCRF Bridge Dataset.}
    \begin{tabular}{c|c|c|c|c}
    \hline  
    Network model & Accuracy & Precision & Recall & F\_1  \\  
    \hline  
    DNN	& $0.929 \pm 0.002$ & $0.918 \pm 0.003$ & $0.928 \pm 0.003$ & $0.925 \pm 0.002$ \\ 
    CNN	& $0.932 \pm 0.004$ & $0.913 \pm 0.005$ & $0.928 \pm 0.004$ & $0.920 \pm 0.004$ \\ 
    LSTM& $0.909\pm0.002$ & $0.893\pm0.003$ & $0.908\pm0.003$ & $0.897\pm0.003$ \\ 
    GRU	& $0.907\pm0.003$ & $0.880\pm0.002$ & $0.923\pm0.003$ & $0.890\pm0.002 $\\   \hline 
    HCG	& \textbf{0.946}$\pm 0.003$ & \textbf{0.920}$\pm 0.003$ & \textbf{0.945}$\pm 0.002$ & \textbf{0.930}$\pm 0.002$ \\ 
    \hline  
    \end{tabular}
\label{tab:hei-acc_loss}    
\end{table}  

\begin{figure}[htbp]
 \centering
 \subfigure{
    \includegraphics[scale=0.15]{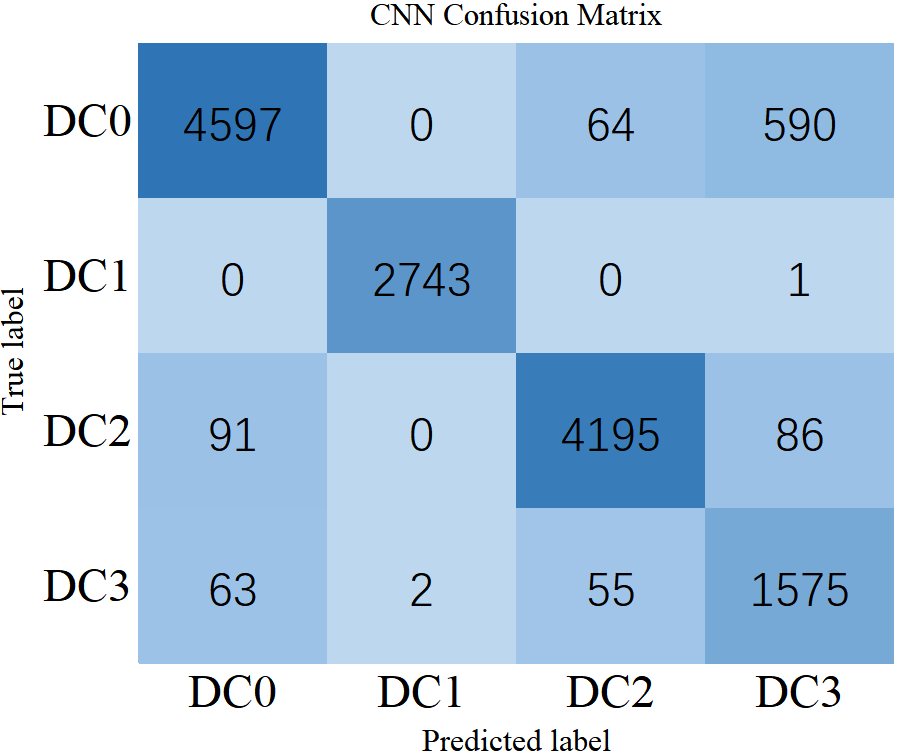}}
 \subfigure{
    \includegraphics[scale=0.15]{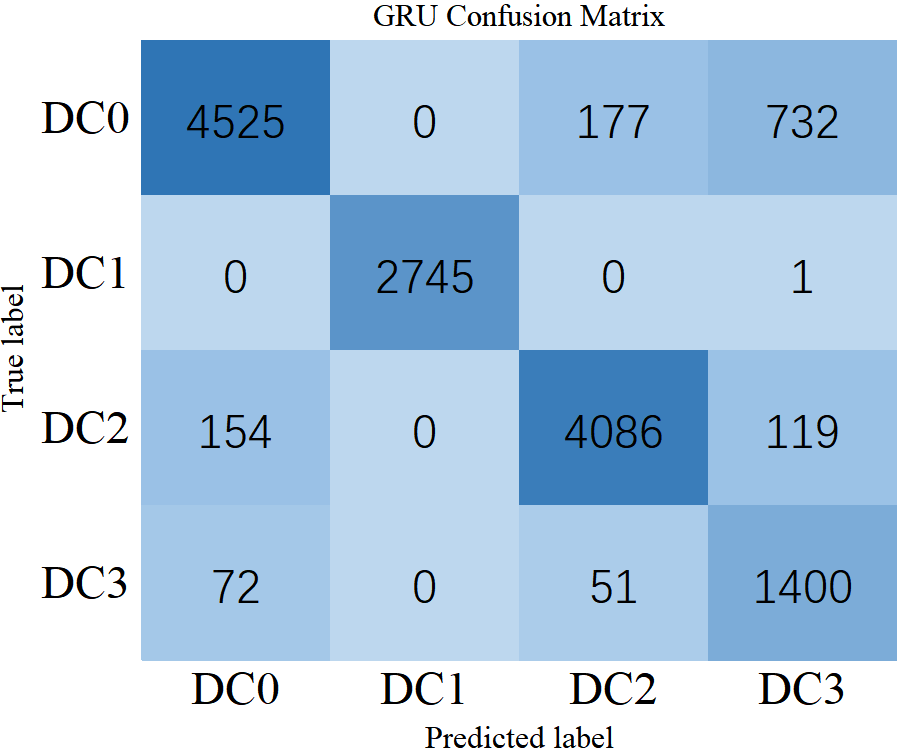}}
 \subfigure{
    \includegraphics[scale=0.15]{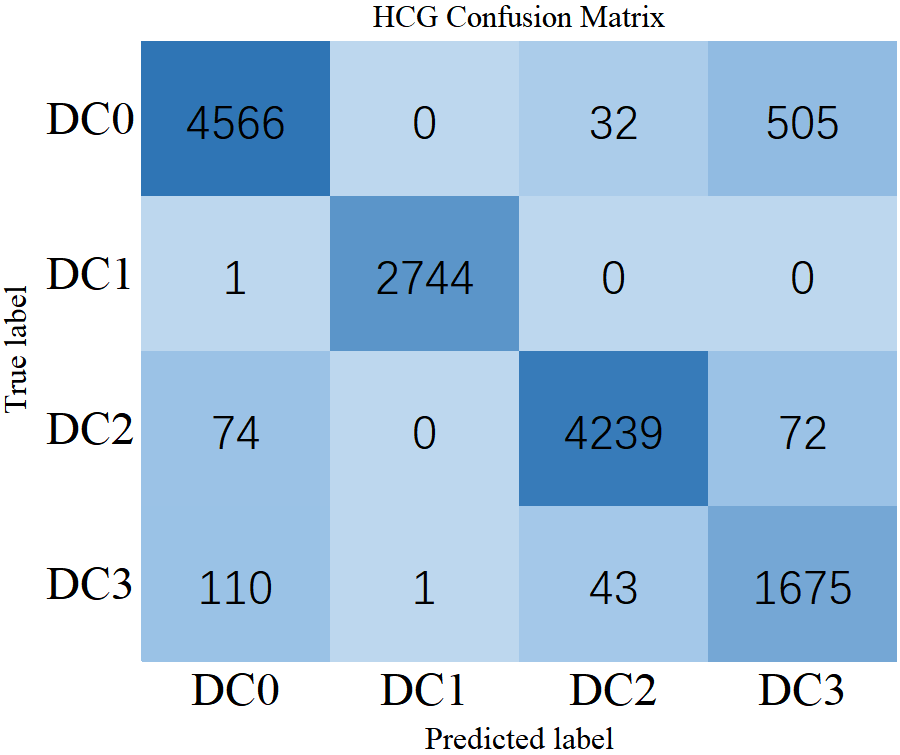}}
\caption{Confusion matrices of CNN based models, GRU based models and our proposed HCG on the TCRF Bridge Dataset.}
\label{hei-hunxiao}
\end{figure}

In order to directly show the ability of the models to identify the damage states, the confusion matrices of CNN, GRU and HCG are presented according to the final results of the test set. As shown in Fig.~\ref{hei-hunxiao}, the HCG model proposed in this paper is used for condition $2$. The classification effect of working condition 3 is higher, while the classification effect of working condition 3 and 4 is general, because the data characteristics of working condition 3 and 4 are similar. The final accuracy of HCG is 94.04\%.

\subsection{Performance Analysis of Hyper-parameters in the TCRF Bridge Dataset}
In order to further demonstrate the advantages of the proposed HCG method rather than a set of a selected hyper-parameters, we compare and analyze the performance of different hyper-parameters, including network structure and number of neurons, of DNN, CNN, LSTM, RNN, and HCG for the TCRF Bridge dataset. 

We first conduct a set of experiments to study the effectiveness of the network structure. We adopt the network structure shown in 
Table \ref{tab:suo11_s} to measure and compare the performance of the DNN, CNN, LSTM, RNN, and HCG model. In the table, $2$-layer, $3$-layer, $4$-layer, $5$-layer present the numbers of neural network layers for the models. The number of neurons is the same for all the models for fair comparisons.  

% Number of neurons for each layer in 2-layer neural network models are $40$, and $40$ respectively, for the 3-layer models are $40$,  $40$, and $70$ respectively, for the 4-layer models are  $40$,  $40$, $70$ and $70$ respectively, and for the 5-layer models are $40$,  $40$, $70$, $70$ and $100$ respectively.

\begin{table}[htpb]
\centering
\caption{Results on different layers for the TCRF Bridge dataset} %%%
\begin{tabular}{c|c|c|c|c}
\hline
Model & 2-layer & 3-layer & 4-layer & 5-layer \\ \hline
DNN   & $0.925\pm 0.001$  & $0.929\pm 0.002$  & $0.934 \pm 0.002$  & $0.933\pm 0.002$  \\
CNN   & $0.928\pm 0.002$  & $0.932\pm 0.002$  & $0.934 \pm 0.002$  & $0.931 \pm 0.001$  \\
LSTM  & $0.863\pm 0.002$  & $0.909\pm 0.001$  & $0.902\pm 0.003$  & $0.882\pm 0.003$  \\
GRU   & $0.892\pm 0.002$  & $0.907\pm 0.003$  & $0.900\pm 0.003$  & $0.887\pm 0.002$  \\\hline
HCG   & \textbf{0.945}$\pm 0.002$  & \textbf{0.946}$\pm 0.002$  & \textbf{0.948}$\pm 0.002$  & \textbf{0.947}$\pm 0.001$  \\ \hline
\end{tabular}
\label{tab:suo11_s}
\end{table}

We then conduct another set of experiments to study the effectiveness of the number of neurons. We adopt the $4$-layer set, because the $4$-layer set achieves higher accuracy than others. We choose different numbers of neurons for all the models. The results are listed in Table~\ref{tab:suo11_p}. In the table, ${[}40, 70, 32, 32{]}$ means $40$, $70$, $32$, and $32$ neurons for the $4$ layers, respectively.

\begin{table}[htpb]
\centering
\caption{Results on different numbers of neurons for the TCRF Bridge dataset}%%%%
\begin{tabular}{c|c|c|c|c}
\hline
Model & {[}40, 70, 32, 32{]} & {[}40, 70, 32, 64{]} & {[}40, 70, 64, 64{]} & {[}40, 70, 64, 100{]} \\ \hline
DNN   & $0.924\pm 0.003$       & $0.928\pm 0.001$      & $0.933\pm 0.0.3$      & $0.932\pm 0.001$         \\
CNN   & $0.927\pm0.002$       & $0.932\pm 0.004$      & $0.935\pm 0.002$      & $0.933\pm 0.002$      \\
LSTM  & $0.864\pm 0.002$       & $0.907\pm 0.002$      & $0.904\pm 0.003$      & $0.881\pm 0.002$       \\
GRU   & $0.892\pm 0.003$       & $0.905\pm 0.003$      & $0.902\pm 0.003$      & $0.888\pm 0.001$      \\ \hline
HCG   & $\textbf{0.9469}\pm 0.003$  &  $\textbf{0.944}\pm 0.002$      &$ \textbf{0.946}\pm 0.003$      & $\textbf{0.949}\pm 0.002$      \\  \hline
\end{tabular}
\label{tab:suo11_p}
\end{table}
From both the tables, it can be concluded that HCG has higher accuracy in different hyper-parameter sets compared with neural network model. HCG model learns from interactions among sensors and the short-term temporal dependencies, and a high-level component that handles
information across long-term temporal dependencies, which has excellent advantages.
%-----------------------------
\subsection{Results of the IASC-ASCE Benchmark Dataset}	
% As shown in Figure ~\ref{benchmark-acc_loss}, the accuracy and loss rate of CNN, DNN, GRU, LSTM and HCG neural network models vary with the number of iterations in the Benchmark finite element simulation dataset. Table ~\ref{tab:Structural} shows the Accuracy and Loss of each model in the validation set.

\begin{figure}[htbp]
  \centering
  \subfigure{
    \includegraphics[width=5.8cm]{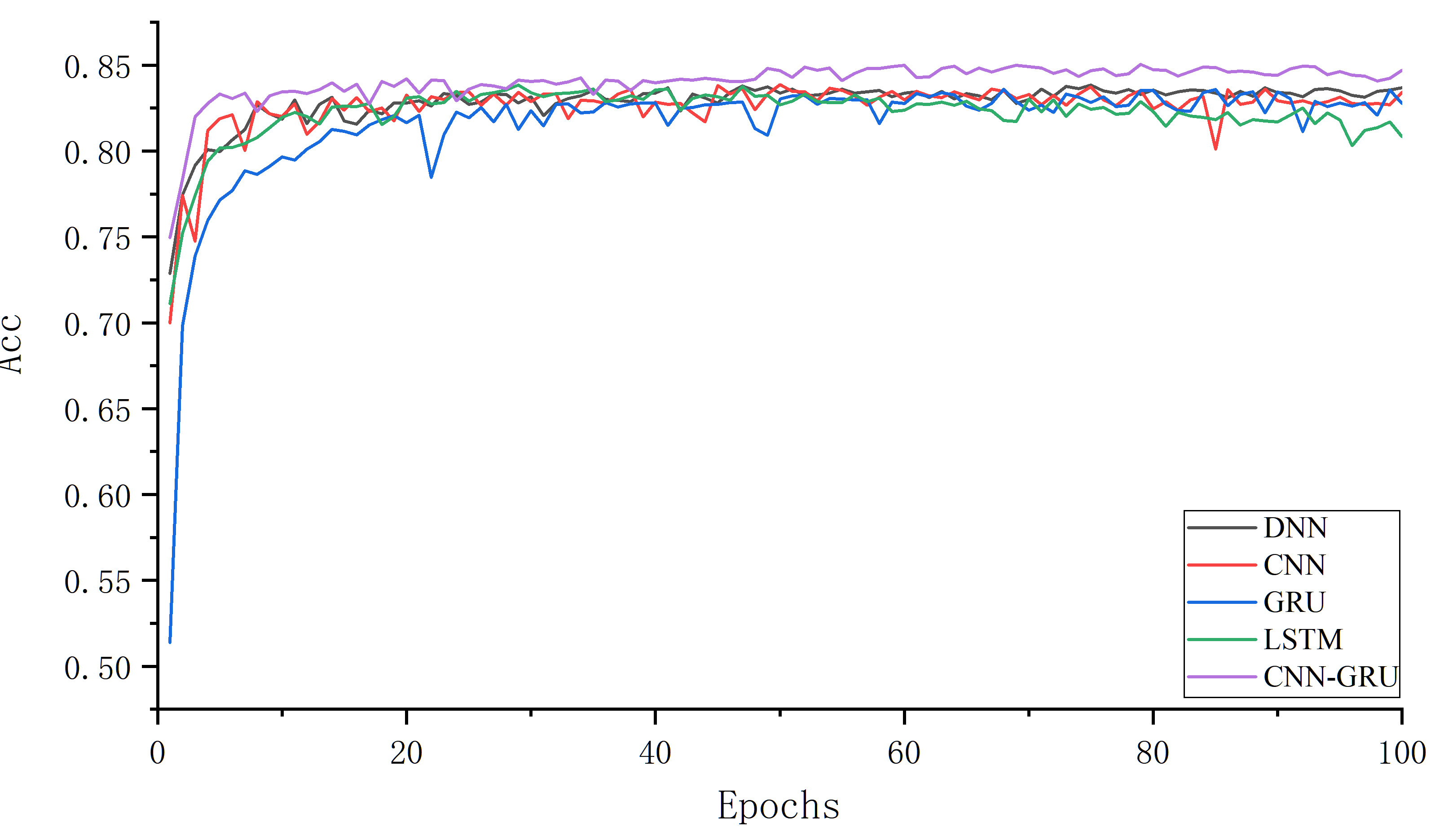}}
  \subfigure{
    \includegraphics[width=5.8cm]{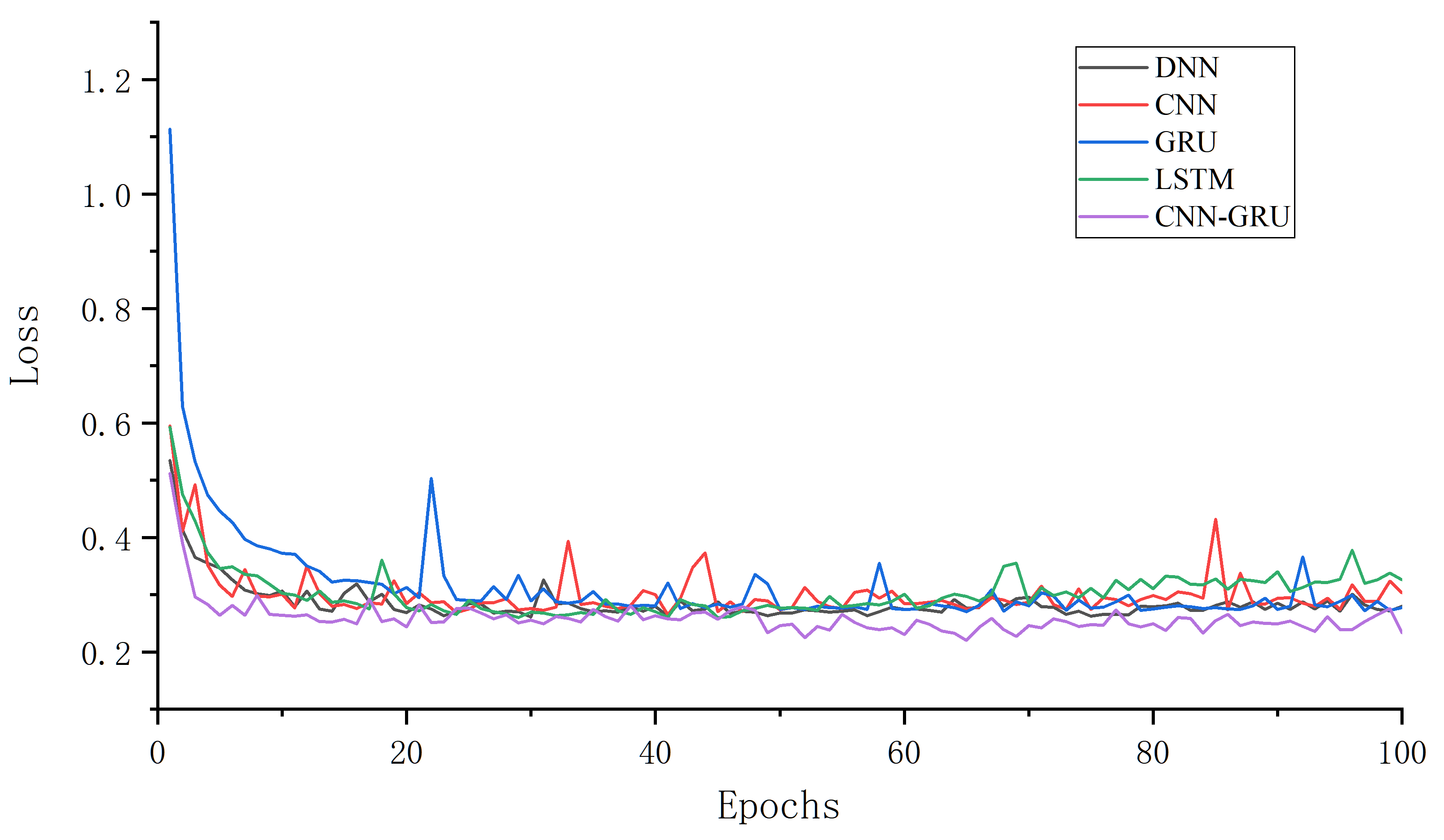}}
\caption{Loss curve and accuracy curve in the training process for the IASC-ASCE Benchmark Dataset. (a) The accuracy curve. (b) The loss curve.}
\label{benchmark-acc_loss}
\end{figure}

Table \ref{tab:Structural} summarizes the experimental results on the IASC-ASCE Benchmark Dataset. Fig.~\ref{benchmark-acc_loss} illustrates the loss curve and accuracy curve in the training process for the deep-learning based baselines and our HCG. Similar with the result in the IASC-ASCE Benchmark Dataset, HCG also shows its advantages over other compared baselines.

\begin{table}[htbp]
    \centering
    \caption{Results of the IASC-ASCE Benchmark Dataset.}
    \begin{tabular}{c|c|c|c|c}
    \hline  
    Network model & Accuracy &Precision &Recall & F1-score \\  
    \hline  
    DNN & $0.837\pm0.002$ & $0.837\pm0.002$ & $0.867\pm0.001$ & $0.768\pm0.001$ \\ 
    CNN	& $0.836\pm0.003$ & $0.836\pm0.001$ & $0.890\pm0.003$ & $0.771\pm0.003$ \\ 
    LSTM& $0.833\pm0.002$ & $0.834\pm0.002$ & $0.887\pm0.002$ & $0.768\pm0.003$ \\ 
    GRU & $0.834\pm0.003$ & $0.834\pm0.003$ & $0.881\pm0.003$ & $0.764\pm0.002$ \\   \hline
    HCG	& $0.841\pm0.001$ & $0.841\pm0.002$ & $0.909\pm0.002$ & $0.781\pm0.001$ \\ 
    \hline  
    \end{tabular}
\label{tab:Structural}
\end{table}  

\subsection{Performance Analysis of Hyper-parameters in the IASC-ASCE Benchmark Dataset}
% Comparing different structure and parameters of DNN, CNN, LSTM, RNN, HCG neural network model. We adopt grid research means to gain  the optimal structure and parameters of the model. The part of result of these model is shown as follow Table \ref{Tab:bench_db}, which conclude that HCG have higher accuracy.

Like in TCRF Bridge Dataset, in order to further demonstrate the advantages of the proposed HCG method rather than a set of a selected hyper-parameters, we compare and analyze the performance of different hyper-parameters, including network structure and number of neurons, of DNN, CNN, LSTM, RNN, and HCG for the IASC-ASCE Benchmark Dataset. 

We first conduct a set of experiments to study the effectiveness of the network structure. We adopt the network structure as shown in Table \ref{tab:ben_s1} to measure and compare the performance of the DNN, CNN, LSTM, RNN, and HCG model. In the table, $2$-layer, $3$-layer, $4$-layer, $5$-layer mean the numbers of neural network layers for the models. The number of neurons is the same for all the models for fair comparisons.

% Parameters of 2layer neural network model are  $32$,  $32$ respectively. Parameters of 3layer neural network model are  $32$,  $32$, $48$ respectively. Parameters of 4layer neural network model are  $32$, $32$, $48$，$48$respectively. Parameters of 5layer neural network model are  $32$, $32$, $48$, $48$, $60$ respectively. These contents are listed in Table ~\ref{tab:ben_s1}.

\begin{table}[htbp]
\caption{Results on different layers for the IASC-ASCE Benchmark Dataset}%%%%
\centering
\begin{tabular}{c|c|c|c|c}
\hline
Model & 2-layer & 3-layer & 4-layer & 5-layer \\ \hline
DNN   & $0.836\pm0.004$  & $0.837\pm0.001$  & $0.839\pm0.002$ & $0.831\pm0.002$  \\
CNN   & $0.832\pm0.005$  & $0.836\pm0.002$  & $0.840\pm0.003$ & $0.836\pm0.001$  \\
LSTM  & $0.835\pm0.002$  & $0.833\pm0.003$  & $0.831\pm0.002$ & $0.832\pm0.002$  \\
GRU   & $0.837\pm0.002$  & $0.834\pm0.001$  & $0.839\pm0.002$ & $0.833\pm0.004$  \\\hline
HCG   & $\textbf{0.841}\pm0.002$  & $\textbf{0.841}\pm0.002$  & $\textbf{0.842}\pm0.003$  & $\textbf{0.842}\pm0.003$  \\ \hline
\end{tabular}
\label{tab:ben_s1}
\end{table}

We then conduct another set of experiments to study the effectiveness of the number of neurons. We adopt the $4$-layer set because the $4$-layer set achieves higher accuracy than others. We choose different numbers of neurons for all the models. The results are listed in Table ~\ref{tab:ben_p}.

\begin{table}[htbp]
\centering
\caption{Results on different numbers of neurons for the IASC-ASCE Benchmark Dataset}%%%%
\begin{tabular}{c|c|c|c|c}
\hline
Model & {[}32, 32, 32, 64{]} & {[}32, 32, 64, 64{]} & {[}32, 64, 64, 64{]} & {[}64, 64, 64, 64{]} \\ \hline
DNN   & $0.835\pm0.002$            & $0.836\pm0.003$            & $0.839\pm0.002$            & $0.831\pm0.002$            \\
CNN   & $0.834\pm0.002$            & $0.834\pm0.002$            & $0.840\pm0.001$            & $0.831\pm0.002$            \\
LSTM  & $0.833\pm0.004$            & $0.834\pm0.001$            & $0.835\pm0.002$            & $0.834\pm0.003$           \\
GRU   & $0.839\pm0.003$            & $0.833\pm0.002$           & $0.839\pm0.002$            & $0.832\pm0.004$            \\\hline
HCG   &   $\textbf{0.841}\pm0.002$            &   $\textbf{0.843}\pm0.002$            &   $\textbf{0.842}\pm0.002$            &   $\textbf{0.843}\pm0.002$            \\ \hline
\end{tabular}
\label{tab:ben_p}
\end{table}
From both the tables, it can be concluded that HCG has higher accuracy in different hyper-parameter sets compared with neural network models.
%--------------------
\subsection{Running Time and Model Sizes}
Table \ref{tab:training} summarizes the running time and the model sizes of different methods. 
We report the running time to finish one epoch
training for all the deep learning models and observe that models
adopted with HCG are roughly $1.5$ times faster than GRU-based
models, which demonstrates HCG is generally much
faster than GRU. Moreover, adding the hierarchical structure in
HCG only slightly affects the computational speed.

GPU memory consumption consists of the memory to store model parameters, which is the same
for all the models in our experiments, and the memory to store the
input data. 
From Table \ref{tab:training}, we can observe that our proposed HCG has the smallest model sizes compared with CNN, LSTM, and GRU.

\begin{table}[htbp]
\caption{Comparison results of the time to finish one epoch training and the model sizes. }
\normalsize
\label{tab:training}
\centering
        \begin{tabular}{c|c|c|c}
\hline
		{Datasets} & {Methods} &  {Training} & {Model Sizes} \\
\hline
\multirow{4}{*}{TCRF Bridge} 
               &{CNN}  & {{32s}} & {{6263KB}} \\
               &{LSTM}  & {{84s}} & {{815KB}} \\
               &{GRU}  & {{80s}} & {{747KB}} \\
               &{HCG}  & {{52s}} & {{721KB}} \\
\hline
\multirow{4}{*}{IASC-ASCE Benchmark} 
               &{CNN}  & {{34s}} & {{4732KB}} \\
               &{LSTM}  & {{75s}} & {{817KB}} \\
               &{GRU}  & {{78s}} & {{750KB}} \\
               &{HCG}  & {{50s}} & {{747KB}} \\
\hline
\end{tabular}
\end{table}

%% file: conclusions.tex
\section{Conclusions}\label{sec:conclusion}
Structural damage detection has become an interdisciplinary area of interest for various engineering fields. In this work, we propose a novel hierarchical deep CNN and GRU framework, termed as HCG, for structural damage detection. HCG is prominent in capturing both the spatial and temporal features among the sensory data. In HCG, we propose a low-level convolutional component to learn the spatial and short-term temporal features. Then, we propose a high-level recurrent component to learn the long-term temporal dependencies. We have done extensive experiments on ASCE Benchmark and a three-span continuous rigid frame bridge structure datasets. The experimental results demonstrate that our HCG outperforms other methods for structural damage detection. To further improve the efficiency of computing, advanced machine learning techniques, such as stochastic configuration networks~\cite{lu2020ensemble, wang2017stochastic,wang2017stochasticFA} and graph convolutional neural network~\cite{guo2019attention}, would be considered in the future research work.